\documentclass [11pt]{article}
\usepackage{todonotes}
\usepackage{hyperref}
\usepackage{listings}
\usepackage{rotating}
\usepackage{makecell}
\usepackage{graphicx}
\usepackage{tabularx}
\usepackage{ragged2e}
\usepackage{caption}
\usepackage{longtable}
\usepackage{booktabs}
\usepackage{array}
\usepackage[backend=biber,style=ieee]{biblatex}
\title{Symbolic Execution in Practice: A Survey of Applications in Vulnerability, Malware, Firmware, and Protocol Analysis}
\author{Joshua Bailey (jbailey4@umbc.edu)\\Charles Nicholas (nicholas@umbc.edu)\\University of Maryland, Baltimore County}

\begin{document}
\maketitle
\section{Introduction}
\label{sec:introduction}
Program testing is an essential aspect of software development. Testing not only helps to verify a program's capabilities, it also uncovers potential vulnerabilities that adversaries could exploit, and offers an opportunity to verify the correctness of an implementation. Traditionally, program verification relied on rigorous formal methods to prove the correctness of a program; while testing, it could be said, used a more practical, but less exhaustive approach where inputs were carefully chosen and run against the program. The outputs were manually inspected and compared against expected outcomes~\cite{King76}. Although formal methods provide strong guarantees about program behavior, its often impractical against large scale software. Simpler testing strategies, while fast, may fail to detect deep, subtle bugs. Because of the complexity of modern software, a middle ground that provides both thoroughness and practicality became necessary.

In the 70s, researchers proposed such a middle ground in {\em symbolic execution}~\cite{howden1977symbolic, boyer1975select, King75, King76}. At the time, the approach was limited by the capability of theorem provers. However, the advent and advancement of satisfiability modulo theories (SMT) solvers~\cite{moura2008z3, barrett2011cvc4} have since made symbolic execution practical. The core idea is to analyze a program with symbolic variables that represent all possible inputs instead of concrete ones. This allows a symbolic execution engine to systematically explore multiple execution paths, offering greater code coverage and more precise bug finding.

This paper presents a comprehensive survey of how symbolic execution aids in the analysis of large and complex software systems. Rather than simply cataloging tools and techniques, we synthesize the surveyed literature to reveal a unifying theme: strategies to manage the complexity of symbolic execution is critical to practicality. We introduce a framework that categorizes these techniques into broad categories: scope reduction, hybrid analysis, and a taxonomy of guidance heuristics. We use this framework to analyze applications across diverse domains. The low-level mechanics of search strategies have been previously surveyed~\cite{sabbaghi2020systematic}, our work provides a higher-level analysis of how these strategies are applied in practice in domains such as vulnerability research, program verification, emulation and firmware analysis, and obfuscated and malicious code analysis. For an introduction to many of these domains, we refer the reader to a modern software engineering textbook such as \cite{SWEBOK2014}.

By examining representative tools and relevant research, we aim to help researchers and practitioners understand when and why symbolic execution is beneficial, and how they can select or adapt tools for their specific scenarios. In doing so, we highlight current strengths and identify areas requiring further development.

The remainder of this paper is organized as follows. Section 2 provides an overview of symbolic execution, detailing the core challenges and common mitigations in the field. Section 3 presents a landscape of popular symbolic execution engines, examining their high-level design, strengths and weaknesses. Section 4 introduces a taxonomy of guidance heuristics commonly employed in symbolic execution research and surveys how symbolic execution has been applied in various application domains. Section 5 outlines potential research directions. Section 6 offers concluding thoughts.

\section{Symbolic Execution Fundamentals}
\label{subsec:symbolic-execution-fundamentals}

Symbolic execution is a program analysis technique that explores executable paths by treating inputs as symbolic variables. For example, instead of running the program with specific inputs like \texttt{x = 4}, symbolic execution uses symbolic values, like \texttt{x = }$\lambda$,  to represent a range of possible inputs, generating expressions that capture the behavior of the program along different paths. 

As the symbolic execution engine traverses a program, it builds up \textit{path conditions} which are logical expressions representing the constraints on inputs required to follow a particular execution path. In addition, the symbolic execution engine may also maintain a store of symbolic memory which maps variables to symbolic expressions or values. In order to determine whether an execution path is feasible, the engine leverages a \textit{constraint solver} (e.g., SMT solvers like Z3 or CVC4) which act as the brains of the operation. Path conditions are fed to the constraint solver and, if a feasible path exists, it will generate concrete inputs that trigger specific behaviors or bugs.

Symbolic execution allows for:

\begin{itemize}
    \item \textbf{Comprehensive Path Exploration}: By considering multiple execution paths, symbolic execution can uncover edge cases and hidden bugs that might be missed by traditional testing. For example, in a program that computes the square root of a real number, symbolic execution should introduce conditions where the input is negative, zero, or positive, whereas traditional testing might only check positive inputs.
    \item \textbf{Automated Test Case Generation}: Symbolic execution can generate inputs that achieve high code coverage, aiding in thorough software testing. 
    \item \textbf{Vulnerability Detection}: Symbolic execution can identify security vulnerabilities by exploring paths that lead to unsafe states or violate security properties.
\end{itemize}

Despite its effectiveness and versatility, symbolic execution is not without its challenges. Limitations such as path explosion, constraint solving, and scalability must be carefully evaluated to determine its feasibility for a given project.

\subsection{Common Challenges in Symbolic Execution}
\label{subsec:common-challenges-in-symbolic-execution}

As noted above, symbolic execution is a powerful technique program analysis technique, but it is not without its difficulties. The core challenges include path explosion, constraint solving difficulties,  and environment modeling. In this section, we will describe these challenges, and others, in greater detail. Later, in Section~\ref{subsec:advancements-in-symbolic-execution} we discuss ongoing research and advances aimed at addressing these issues.

\subsubsection{Path Explosion}
\label{subsubsec:path-explosion}
Symbolic execution's strength lies in its ability to explore all possible executions paths of a program simultaneously. While this property allows for thorough code coverage, it's this very strength that's also one its greatest weaknesses. The sheer number of paths to analyze can make analysis increasingly resource intensive. This phenomenon is known as path or state space explosion. As the number of paths increases, the symbolic execution engine has to manage memory, CPU time, and solver queries, ultimately limiting the scalability of the analysis. The following examples, while trivial, illustrate how quickly this complexity can arise.

A common cause of path explosion are {\em loops with symbolic conditions}. Consider the code snippet in Listing~\ref{lst:path-explosion-loop}:

\begin{lstlisting}[language=c,caption={A loop with a symbolic condition}, label={lst:path-explosion-loop}]
    void loop_example(int n) {
        for (int i = 0; i < n; i++) {
            // Some code dependent on 'i'
        }
    }
\end{lstlisting}
In Listing~\ref{lst:path-explosion-loop}, if \texttt{n} is symbolic, the loop's iteration count is effectively unbounded. The symbolic execution engine must consider all possible values of \texttt{n}, leading to a potentially infinite number of paths. To mitigate infinite exploration, symbolic execution engines typically impose loop unrolling limits~\cite{jaffar2012unbounded}. Alternatively, the variable could be concretized or constrained. However, such heuristics risk missing bugs that only manifest at larger iteration counts. Techniques like loop summarization~\cite{godefroid2011automatic} or state merging~\cite{kuznetsov2012efficient} can help reduce the number of paths while preserving a meaningfully high level of coverage. 

{\em Nested conditionals} are another major contributor to path explosion. Listing~\ref{lst:path-explosion-nested} shows how even a small number of symbolic conditionals can cause exponential growth of the state space:

\begin{lstlisting}[language=c, caption={Path explosion caused by nested conditional statements}, label={lst:path-explosion-nested}]
    void conditional_example(int a, int b) {
        if (a < 0) {
            if (b > 0) {
                // Path 1
            } else {
                // Path 2
            }
        } else {
            if (b > 0) {
                // Path 3
            } else {
                // Path 4
            }
        }
    }
\end{lstlisting}
For two symbolic variables \texttt{a} and \texttt{b}, there are four distinct paths. Adding a third symbolic variable \texttt{c} would create eight paths. Because symbolic execution must analyze the true and false branch every time a conditional expression is encountered. Consequently, the symbolic execution engine increases the number of paths by a factor of two, leading to an exponential growth of up to $2^n$ paths. 

Certain design decisions, like {\em input-dependent branching}, exacerbate the path explosion problem. Consider the code in Listing ~\ref{lst:path-explosion-input} that independently checks multiple input bytes.  Here, the variable \texttt{input} is treated as a symbolic array where each element represents a symbolic variable. While structurally different from nested conditionals, this pattern produces the same exponential effect since each \texttt{if} statement also introduces two branches. This branching pattern appears frequently in device drivers, file parsers, and embedded systems.
\begin{lstlisting}[language=c,caption={Path explosion caused by input-dependent branching}, label={lst:path-explosion-input}]
    void input_branching(char *input) {
        if (input[0] == 'A') {
            // case where first character is 'A'
        }
        if (input[1] == 'B') {
            // case where second character is 'B'
        }
        if (input[2] == 'C') {
            // case where third character is 'C'
        }
        // ... additional conditions
    }
\end{lstlisting}
 In each of these examples, the engine must carefully track path conditions and program state which can become unwieldy and consume substantial resources. Techniques such as path pruning and heuristic search strategies can tame path explosion, though each comes with their own limitations, discussed in Section ~\ref{subsec:advancements-in-symbolic-execution}. 

\subsubsection{Constraint Solving Difficulties}
\label{subsubsec:constraint-solving-difficulties}
Constraint satisfaction problems are applicable in a wide array of applications such as software and hardware verification, type-checking, test-case generation scheduling, planning, to name a few~\cite{de2009satisfiability}. Boolean satisfiability (SAT) is a well-known constraint satisfaction problem that determines if a formula over Boolean variables can be made true. SAT solvers, while powerful, are limiting in that they are not able to reason about complex logic that might be encountered in modern programming languages. Satisfiability Modulo Theories (SMT) generalize the SAT problem to make them more amenable to such complex paradigms such as bit-vectors, arrays, and linear integer and real arithmetic, and non-linear arithmetic. Consider the simple function \texttt{check}~\ref{lst:check-function}:
\begin{lstlisting}[language=c,caption={An illustration of path condition generation}, label={lst:check-function}]
    void check(int x) {
        int y = x * 2;
        if (y == 12) {
            // Interesting path
        }
    }
\end{lstlisting}
If we want to determine whether the interesting path is reachable, the symbolic execution engine would generate the following path condition \texttt{(x\_sym * 2) == 12}, where \texttt{x\_sym} is a symbolic variable. This formula is sent to the SMT solver where it's task is to determine if a value exists that makes this formula true. In this example, the SMT solver would determine this path is satisfiable and provide the concrete example, \texttt{x = 6}.

Symbolic execution heavily relies on constraint solvers to determine path feasibility. By only analyzing feasible paths, symbolic execution can accurately explore multiple execution paths and reason about program behaviors and variable properties. Early symbolic execution work by King~\cite{King76} demonstrated that constraint solving is fundamental since symbolic execution requires evaluating path conditions and determining satisfiability of symbolic expressions. However, King noted that the theorem proving capabilities available at the time made symbolic execution impossible even for modest programming languages. At that time, efficient theorem solvers capable of handling the complex constraints generated during symbolic execution were not available.%, limiting its practical application. 
However, modern SMT solvers like STP~\cite{ganesh2007Decision} and Z3~\cite{moura2008z3} have enabled practical analysis by efficiently handling complex constraints over arithmetic, bit-vectors, arrays, and other operations.
% citing these papers again is okay, since the first citation was some pages back.  As a rule of thumb, each distinct quote or idea from a given paper can be given its own citation.
Despite these advancements, constraint solving remains a bottleneck. Certain classes of constraints, such as non-linear arithmetic constraints ~\cite{abraham2017smt}, remain difficult to solve, and the presence of path explosion and complex theories can limit scalability. Even powerful solvers such as Z3 cannot fully mitigate these challenges, requiring researchers and practitioners to employ heuristics, approximations, or domain-specific optimizations.

\subsubsection{Environmental Interaction Modeling}
\label{subsubsec:environmental-interaction-modeling}
Any useful software application does not operate in a vacuum. Even modest application must interact with their environment via I/O operations, network communication, interfacing with hardware peripherals or some other external component. The challenge for symbolic execution engines is determining how to symbolically represent these interactions. As a result, the symbolic execution engine must simulate file systems, protocols, and devices with sufficient fidelity to avoid unsound assumptions or missing crucial paths. As mentioned in Section~\ref{subsubsec:multi-threading-and-concurrency}, concurrency can exacerbate environmental modeling complexity, as timing-dependent interactions across multiple threads interacting with external systems increase the number of scenarios to consider.
Domain-specific knowledge, auxiliary models, or hybrid symbolic-concrete execution strategies are often required. However, these solutions may limit general applicability or require extensive manual effort.

Understanding these core challenges: path explosion, floating-point handling, concurrency, constraint solving difficulties, obfuscated code, scalability, and environmental modeling, provides a foundation for evaluating symbolic execution tools and the research efforts which
aim to overcome these hurdles. In Section~\ref{subsec:advancements-in-symbolic-execution}, we explore ongoing efforts and innovative techniques designed to address these persistent issues.

\subsubsection{Other Challenges}
Beyond the core foundational issues, there are more specialized issues that hamper the technique's effectiveness. 

\paragraph{Floating-Point Arithmetic Handling}
\label{subsubsec:floating-point-arithmetic-handling}
Floating-point arithmetic is a critical part of modern computing, specifically in scientific computing where accuracy is paramount. However, real numbers are infinite so representing them in a finite binary format pose a real challenge in computer systems. While the IEEE 754~\cite{ieee-floating-point1985} standard provides a uniform representation of floating-point values, it does not eliminate the fundamental rounding errors and precision loss that can make program behavior difficult to reason about~\cite{goldberg1991every}.

For symbolic execution, handling floating-point arithmetic is especially challenging. The issues primarily stem from constraint solvers having difficulty solving formulas with floating-point values. Many solvers handle floating-point constraints using approximations, potentially leading to incomplete or inaccurate analyses~\cite{liew2017floating}. 

As a result, floating-point computations remain an open problem area, with ongoing research trying to provide sound and efficient handling~\cite{liew2017floating, fu2017achieving}. For a comprehensive survey of the different techniques being developed to address these challenges, we refer the reader to work by Zhang  et al.~\cite{zhang2022symbolic}. 

\paragraph{Multi-threading and Concurrency}
\label{subsubsec:multi-threading-and-concurrency}
Multi-threaded programs enable software to make efficient use of CPU resources. Threaded applications are able to handle multiple streams of input concurrently, perform complex computations in parallel, and distribute work among various compute resources. While threads offer significant performance gains, it also introduces substantial challenges in debugging and analysis due to the inherent complexity and non-determinism of concurrent execution.

The non-determinism arises because the order in which threads execute is not strictly defined. Consequently, certain bugs are difficult to reproduce since the same sequence of operations might not occur consistently across multiple runs. Furthermore, multiple threads accessing the same shared resource without proper synchronization can lead to race conditions causing unpredictable and subtle-to-detect bugs. Another consequence of threads is the exacerbation of the path explosion problem. Moreover, the shared state must be accurately modeled across all thread, significantly increasing the complexity of the analysis. As discussed in Section~\ref{subsubsec:environmental-interaction-modeling}, concurrency can also magnify the challenges of modeling external environments. Timing and ordering constraints introduced by multiple threads interacting with external resources or devices can greatly complicate environmental modeling. This forces the symbolic execution engine to consider more nuanced scenarios.

Approaches to mitigating concurrency-related challenges, such as program sequentialization~\cite{bakst2017verifying}, can simplify thread interleaving but introduce scalability issues, risk losing concurrency semantics, and rely on simplifying assumptions that may overlook critical issues.

\paragraph{Handling Obfuscated and Self-Modifying Code}
\label{subsubsec:handling-obfuscated-and-self-modifying-code}
Code obfuscation is a technique that transforms a program into a form that is more difficult to understand. Obfuscation techniques such as self-modifying code complicate analysis by altering control flow and data structures in ways that hinder the engine's assumptions and simplifications. Self-modifying code forces a symbolic execution engine to adapt dynamically to changing code and maintain soundness despite transformations intended to confuse analysis. To this end, researchers have developed specialized techniques to counter specific types of obfuscation. For instance, some approaches combine symbolic execution with taint analysis to reverse virtualized code~\cite{Salwan2018symbolic}, while others use novel methods like backward-bounded dynamic symbolic execution (DSE)~\cite{bardin2017backward} to defeat opaque predicates. Despite these advances, such solutions are often tailored to particular obfuscation schemes, and creating a robust, general-purpose deobfuscation framework remains a significant and open research challenge. 
%\gripe{can we cite any of this recent work? if not, better to say it's still an open problem.}

\paragraph{Scalability}
\label{subsubsec:scalability}
It is often desirable to analyze software at scale whether it be analyzing one project with millions of lines or analyzing multiple projects efficiently. Applying symbolic execution to large-scale, real-world software systems involves addressing thousands if not millions of lines of code, external libraries, complex data types, and intricate state spaces. The core challenges presented earlier in this section contribute to the challenge of making symbolic execution scalable. Managing resource consumption, which may include efficient path pruning, leveraging parallelization, and distributed analysis, is essential to making symbolic execution practical at scale. Ultimately, achieving scalable analysis with symbolic execution requires mitigating one or more of the challenges presented here. We will discuss various scalability-improving techniques in Section~\ref{subsec:advancements-in-symbolic-execution}.

\subsection{Addressing Core Symbolic Execution Challenges}
\label{subsec:advancements-in-symbolic-execution}
While Section~\ref{subsec:common-challenges-in-symbolic-execution} identified several ongoing difficulties in applying symbolic execution to real-world software, most research efforts have focused on the most pressing and foundational problems, particularly path explosion and constraint solving difficulties. By making path exploration more tractable, these approaches also improve scalability, reduce server load, and indirectly aid in handling other challenges like environmental modeling and complexity in multi-threaded scenarios. In symbolic execution, every path analyzed by the engine must maintain a consistent view of the environment (e.g., registers, file systems, etc.). Reducing the sheer number of states to track, whether it be by pruning or merging paths, reduces the number of environmental states the engine must manage.

In this section, we highlight notable strategies and techniques that have proven effective in mitigating path explosion and, to a lesser extent, improving constraint handling. Although not every challenge from Section~\ref{subsec:common-challenges-in-symbolic-execution} is directly addressed here, many of the solutions are broad in scope, offering partial relief or foundational improvements that can be adapted or extended to tackle additional issues.

\subsubsection{Techniques for Managing Path Explosion and Scalability}
\label{subsubsec:manage-path-explosion}
\paragraph{State Merging and Pruning.\label{sec:state-merging}}

As highlighted in Section ~\ref{subsubsec:path-explosion}, path explosion often occurs because each symbolic branch creates new states. However, not all branches lead to unique states. Therefore similar states can be merged and redundant paths eliminated drastically reducing the number of paths to explore~\cite{kuznetsov2012efficient}. Likewise, certain paths may not yield promising results. These "low-value" paths can be pruned using heuristics to focus analysis on the most promising execution paths~\cite{burnim2008heuristics}. Bounded exploration~\cite{siddiqui2012scaling}, which limits the depth or number of paths explored, further controls resource usage. While these techniques do not solve path explosion entirely, they make symbolic execution more feasible for larger and more complex programs.

\paragraph{Concolic Execution}
\label{sec:concolic-execution}
As we've noted, constraint solving is computationally expensive and remains the primary bottleneck in symbolic execution~\cite{cadar2013symbolic}. To address this, researchers have employed a method that combines concrete execution with symbolic execution resulting in 
what is called {\em concolic} execution. The idea is to utilize concrete runs of the program under test to drive the symbolic execution engine in the right direction. As we will see in Section~\ref{sec:analysis-challenges-and-tool-evaluations} guiding symbolic execution is a popular mechanism to reduce path explosion. Because determining path feasibility is not required (due to the concrete run providing inputs), it reduces the load on the constraint solver which continues to be the largest bottleneck for practical symbolic execution on large programs. In addition, because it is intractable to keep track of the entire environment and effects from a given function call, blending concrete execution with symbolic analysis guides path exploration on the basis of actual execution traces.

\paragraph{Heuristic Search Strategies.\label{sec:heuristic-search}}
Because symbolic execution aims to cover all paths, providing guidance to focus its exploration is critical. As a result, many researchers employ heuristics to prioritize certain paths over others. For example, SAVIOR~\cite{chen2020savior} (detailed in Section~\ref{subsec:vulnerability-research}) uses a bug-driven approach. First, it identifies code regions that trigger a undefined behavior address sanitizer alerts. Then symbolic execution is focused on paths that are more likely to lead to vulnerable code. This selective exploration can rapidly surface interesting program behaviors without exhaustively enumerating all paths, thus mitigating path explosion and accelerating bug discovery.

\subsubsection{Parallelization}
\label{subsubsec:parallelization}
Distributing symbolic execution workloads across multiple processors or machines allows engines to handle more states in parallel. While parallelization does not inherently eliminate path explosion, it leverages additional computational resources to process a larger number of paths simultaneously, improving scalability and practical applicability. One example of this is the Cloud9~\cite{bucur2011parallel} platform which executes symbolic execution work on hardware clusters.
%\gripe{it would be good to name a system that does this}

\subsubsection{Hybrid Approaches}
\label{subsubsec:hybrid-approaches}
As noted earlier, fuzzing alone often yields limited code coverage. In contrast, symbolic execution excels in deeper analysis, but there are severe performance penalties. A promising approach combines both techniques: using a fuzzer for rapid exploration and symbolic execution for precision. The idea is to aid the fuzzer when it becomes "stuck" (i.e., fails to explore new paths), symbolic execution can generate inputs to expand coverage. This synergy can reduce the overall complexity that the symbolic engine faces, indirectly helping with path explosion and constraint-solving challenges. In Section~\ref{sec:analysis-challenges-and-tool-evaluations} there are a few examples of systems which show how fuzzing and symbolic execution can be utilized together. Although hybrid approaches do not directly address all challenges, they improve the engine's efficiency and effectiveness in practice by leveraging symbolic execution when necessary.

Symbolic execution has made significant strides in becoming a viable and practical option for analyzing large-scale software systems, and a wide array of tools exist to support this endeavor. In the next section, we highlight only a few of these tools in an effort to provide a broad overview of the symbolic execution frameworks. Rather than directly comparing the tools against one another, we focus on their diverse applications in tackling challenges such as vulnerability research, program verification, analysis of obfuscated and malicious code, emulation and firmware analysis, and protocol inference. Following Section~\ref{sec:tools-landscape}, we highlight how symbolic execution has been used in various application domains.

\section{Symbolic Execution Tools Landscape}
\label{sec:tools-landscape}
Symbolic execution has matured into a practical technique, supported by a diverse landscape of tools.  While not an exhaustive list, this curated set showcases tools chosen for their influence, current relevance, and varied capabilities. These tools differ in their operational approach: some operate directly on binaries, while others require source code to translate into LLVM bitcode~\cite{lattner2004llvm} and operate on that. 
These tools also differ in their primary focus, supported architectures, and performance characteristics and strategies for mitigating core challenges like path explosion. Understanding these distinctions is crucial for selecting the appropriate tool for a given analysis task.

Symbolic execution tools strive to balance factors such as speed, ease of implementation, and architecture independence. Tools like KLEE~\cite{cadar2008klee} and SymCC~\cite{Poeplau2020symbolic}
%\footnote{Appropriate references will be cited as we begin to describe these systems in detail.}
%\gripe{should cite each tool upon its first mention}
closely integrate with source-level representations (typically LLVM bitcode), enabling comprehensive analysis integrated early in the development cycle. Conversely, tools such as angr and Triton adopt binary-only approaches, lifting machine instructions to an intermediate representation (IR) for analysis.  This lifting is useful in scenarios where source code is not available, such as malware analysis or closed-source vulnerability analysis. Furthermore, frameworks like $S^2E$~\cite{chipounov2011S2E} combine symbolic and concrete execution to handle entire system stacks, including operating systems and device drivers, while hybrid approaches, exemplified by Driller, fuse symbolic execution with fuzzing to leverage the strengths of both techniques.

These diverse approaches can be broadly categorized based on their primary input analysis strategy: source-based analysis, binary-level analysis, and hybrid analysis (often combining symbolic execution with fuzzing or concrete execution). We will now discuss prominent examples within each category, outlining their methodologies, strengths, weaknesses, and typical application domains.

\subsection{Source-Based Analysis}
Source-based tools typically operate on intermediate representations such as LLVM bitcode, requiring access to the original source code or a compilable form. The most prominent tools in this category are KLEE~\cite{cadar2008klee} and SymCC~\cite{Poeplau2020symbolic}.

\paragraph{KLEE}
\label{para:klee}
Developed in 2008 as an evolution of the earlier EXE tool~\cite{cadar2006exe}, KLEE remains one of the most widely used symbolic execution frameworks, cited in numerous publications. KLEE's primary goal is high-coverage automated test generation for C/C++ programs. Key technical contributions include an efficient state representation using object-level copy-on-write memory to minimize state duplication overhead, and optimizations for constraint solving. Specifically, KLEE employs constraint independence (reducing solver queries), significantly reducing the burden on the underlying SMT solver (originally STP, now also supporting Z3). To use KLEE, programs must be compiled to LLVM bitcode. Developers can annotate their code to mark specific inputs as symbolic and add assumptions (e.g., constraining input values). KLEE's reliance on source/bitcode makes it particularly well-suited for integration into software development workflows for thorough testing and bug finding. 

\paragraph{SymCC}
\label{para:symcc}
Introduced in 2020, SymCC~\cite{Poeplau2020symbolic} represents a distinct approach inspired by performance analysis~\cite{Poeplau2019systematic} showing significant overhead from IR interpretation. Instead of interpreting bitcode symbolically like KLEE, SymCC compiles symbolic execution capabilities directly into the binary using a custom compiler pass built on LLVM. While SymCC still requires LLVM bitcode as input, the resulting instrumented binary performs symbolic tracking natively during execution. When run, SymCC's instrumentation interacts with a symbolic backend (runtime library) to explore new paths and generate inputs that increase code coverage. This compile-time approach allows, in theory, for recompiling libraries like libc to achieve comprehensive environment modeling. SymCC also handles calls to uninstrumented code gracefully by treating their effects concretely. The developers provide specialized code symbolic models for common C library functions (e.g., memset and memcpy). While initially focused on C/C++, the authors suggest the approach could be extended to other languages supported by LLVM. 

KLEE and SymCC, while both source-based, reflect different design priorities originating from how they evaluated their tools. KLEE focused on whether test cases generated by KLEE improved code-coverage while SymCC focused on raw speed and correctness.
%\gripe{what evaluations?}
The creators of KLEE primarily emphasized achieving high code coverage demonstrating effectiveness in finding deep bugs in complex software, with constraint solving optimizations aimed at making this feasible. In contrast, the SymCC developers focused heavily on execution performance, comparing runtime overhead of their direct compilation approach against traditional interpretation methods, alongside code coverage metrics. This difference highlights a key trade-off: KLEE's architecture facilitates deep program state analysis and complex environment modeling, while SymCC prioritizes minimizing the performance impact of symbolic instrumentation.
% the two citations in this paragraph seem redundant, so i removed them

\subsection{Binary-Based Analysis}
On the other end of the spectrum there are engines which do not require access to the source code. Due to the lack of semantic information, binary-based tools typically utilize a "lifter" to transform the machine code up to an intermediate representation (IR). The IR used varies, but the results allow for symbolic analysis on programs where source code is not available. The most prominent tools in this category are $S^2E$, \texttt{angr}, Triton, and BINSEC/SE.

\paragraph{\boldmath $S^2E$}
\label{para:s2e}
Introduced in 2011, $S^2E$ (Selective Symbolic Execution Engine)~\cite{chipounov2011S2E} was designed as a general platform for in-vivo multi-path analysis of complex software systems, scaling even to full OS stacks like Windows~\cite{chipounov2011S2E}. It addresses the significant challenge of analyzing software behavior within its real environment (libraries, kernel, drivers) without resorting to potentially inaccurate or labor-intensive abstract models. $S^2E$ achieves this by combining virtualization
(using QEMU~\cite{bellard2005qemu}), dynamic binary translation (interpreting x86 machine code directly), and symbolic execution (leveraging KLEE).

The primary contribution of $S^2E$ is the idea of selective symbolic execution. Instead of executing the entire system symbolically, which is often infeasible due to path explosion, $S^2E$ allows analysts to precisely target specific code regions for symbolic exploration while executing the rest of the system concretely. It manages seamless, automatic, bi-directional transitions between concrete and symbolic modes, ensuring that interactions with the real symbolic state space and focuses analysis effort where needed. $S^2E$ is modular in structure, featuring path selectors (to guide exploration) and path analyzers (plugins that observe or check properties along explored paths). The authors provide several plugins with its release, and they claim that developing new analyzers is trivial. They corroborate this by developing three analyzers using their system, $DDT^+$, a tool for testing closed-source Windows device drivers, $REV^+$, a tool for reverse engineering binary Windows device drivers, and PROFs, a performance profiler and debugger. 

Overall, the strength of $S^2E$ lies in enabling complex, system-wide analyses directly on binaries within their native environment. 

\paragraph{Triton}
\label{parac:triton}
Introduced in 2015, Triton~\cite{saudel2015Triton} is presented as a dynamic binary analysis framework built upon Intel Pin~\cite{reddi2004PIN}. The authors' primary motivation was to provide a modular binary analysis framework that is well suited to handle obfuscation in modern binaries.

Triton offers Python bindings and integrates several key components: dynamic taint analysis ("Hue Engine"), a symbolic execution engine, a snapshot engine for replaying/backtracking execution paths, x86-64 instruction semantics translated to the SMT2-LIB standard format, and an interface to the Z3 SMT solver. These components allow external Python tools built on Triton to perform tasks like symbolic fuzzing, trace analysis, and runtime analysis for vulnerability research. The authors acknowledge path explosion as a general challenge and describe the standard dynamic symbolic execution to negate branch conditions from previous runs. However, they do not explicitly discuss any heuristics to mitigate path explosion beyond the standard dynamic symbolic execution (DSE) workflow.

For constraint solving, Triton performs "backward reconstruction" to build solvable formulas from symbolic expressions. It also provides robust mechanisms to manage constraint complexity through SMT simplification passes. The framework offers two primary approaches to this simplification: an analyst can register custom simplification callbacks based on user-defined rules, or they can convert Triton's internal expressions into a Z3-compatible format to leverage the solver's simplification engine. The authors do not provide a formal evaluation, but rather demonstrate Triton's capabilities through illustrative examples and descriptions of specific runtime analyses implemented using the framework.

\paragraph{angr}
\label{para:angr}
Next to KLEE, \texttt{angr}~\cite{shoshitaishvili2016state}, developed in 2016, is the most influential binary analysis framework. The developers of \texttt{angr} aimed to overcome challenges associated with the reproducibility and comparability of previous research prototypes. Key design goals included cross-architecture support, cross-platform compatibility, support for different analyses, and usability via a Python interface.

In order to achieve cross-architecture support, \texttt{angr} utilizes Valgrind's VEX IR~\cite{nethercote2007valgrind}, although \texttt{angr}'s modular design allows for other IRs. Cross-platform binary loading, including dependencies, is handled by the CLE (CLE Loads Every) module. Program state, including registers, memory, filesystem, etc., is managed by the SimState object within the SimuVEX module, which uses a plugin system to support different memory models and track environment interactions. Just like KLEE, \texttt{angr} has a rich ecosystem of tools built with \texttt{angr} including \texttt{angrop}~\cite{AngropGithub} for automating the construction of ROP chains, \texttt{patchrex}~\cite{PatcherexGithub} for patching binaries, and \texttt{rex}~\cite{RexGithub} for automated exploit generation.

The \texttt{angr} system incorporates several strategies to address common symbolic execution challenges. To address path explosion, \texttt{angr} implements advanced techniques like Veritesting~\cite{10.1145/2568225.2568293} for state merging and support path prioritization strategies. For constraint solving they implement optimizations like splitting constraints into independent sets or using Value-Set Analysis (VSA)~\cite{balakrishnan2010wysinwyx} to approximate solutions quickly. Finally, to model the environment, \texttt{angr} utilizes "SimProcedures" which are Python functions that model their effect on the SimState. 

\paragraph{BINSEC/SE}
\label{para:binsec}
BINSEC/SE~\cite{david2016binsec}, introduced in 2016, is a dynamic symbolic execution toolkit for binary-level security analysis, particularly reverse engineering tasks like malware analysis and vulnerability research. It is based on the open-source BINSEC~\cite{djoudi2015binsec} platform, which provides a formal, modular environment for binary analysis, featuring disassembly, intermediate representation translation, 
simulation, and static analysis capabilities. BINSEC/SE leverages the platform's front-end for translating x86 binaries into the Dynamic Bit-vector Automata (DBA)~\cite{bardin2011bincoa} IR. The BINSEC/SE architecture is notable for its modularity, comprising of a PIN-based tracer (PINSEC), a core DSE engine with advanced constraint optimizations, and a flexible path selection module inspired by OSMOE~\cite{bardin2011osmose}. The initial release of BINSEC/SE only supported x86. However, since its inception, it has evolved to support a broader range of architectures including 64-bit architectures and ARM.

The authors evaluate BINSEC/SE through two reverse engineering case studies:
\begin{itemize}
    \item Solving a Flare-On challenge crackme by using DSE to find the correct input bytes satisfying path conditions, demonstrating the ability to handle iterative constraints and interact with the tracer to force specific paths.
    \item Performing malware exploration on 11 samples from the VX Heaven dataset\cite{UCIVxHeavenDataset}, where BINSEC/SE automatically discovered 43 new behaviors by systematically negating branch conditions found in initial traces.
\end{itemize}

%\gripe{see comments in latex}
% why did you mention that some parts of these systems are written in OCAML?
% take care not to quote too much from the papers you're citing, even though you are citing them properly
% the language should still sound like your (our) words

%i don't like to use "it" when the system name has not been mentioned lately

\subsection{Hybrid Analysis}
\paragraph{Driller}
\label{para:driller}
Driller~\cite{stephens2016driller} is a hybrid vulnerability analysis tool designed to find deep bugs in binary applications by combining fuzzing with concolic execution. Driller aims to address the weaknesses inherent in fuzzing and concolic execution by playing off the strengths of both techniques. Driller attempts to mitigate path explosion by offloading most of the path exploration task to its fuzzing engine, using concolic execution only to satisfy complex checks in the application~\cite{stephens2016driller}.

Driller employs a feedback loop between a fuzzer (American Fuzzy Lop (AFL)~\cite{zalewski_afl}) and \texttt{angr} as its concolic execution engine. The process begins with AFL fuzzing the application. Once AFL fails to find new interesting states, Driller invokes the concolic execution engine which solves for a specific input and passes it back to the fuzzing component.

For the most part, Driller tackles path explosion by relying on the fuzzer for broad path discovery, and invoking the concolic engine only selectively. Likewise, it mitigates the cost of constraint solving by ensuring the solver is only used to generate the specific inputs necessary to bypass checks that are blocking the fuzzer.
%\gripe{still not suure if I understand this last sentence}
Because Driller is built on top of \texttt{angr} and AFL QEMU, it relies on those components for modeling the environment and therefore suffers the same drawbacks.

Driller was evaluated on the DARPA CGC dataset\cite{cgc_datasets} containing 126 binaries. The authors compared it against standalone AFL and pure symbolic execution (\texttt{angr} with Veritesting). Driller did offer some improvements over AFL, identifying crashes in 77 binaries compared to 68 and 16 in AFL and \texttt{angr} respectively.

%\medskip
\subsection{Selecting the Right Tool}
\label{subsec:tool-selection}

While the tools described above offer powerful capabilities, they differ significantly in their underlying methodologies, target scope, and approaches to core challenges. Understanding these distinctions is crucial for researchers to select the appropriate tool for their specific analysis goals and constraints. In a development environment, KLEE and SymCC would be the obvious choices given access to source code. 
Tables~\ref{tab:comparative_overview_symbolic_execution_tools} 
and~\ref{tab:comparative_overview_symbolic_execution_tools2} illustrate the characteristics and use-cases for each of the tools reviewed.

\begin{sidewaysfigure}[p]
\caption{Comparative Overview of Symbolic Execution Tools, Part 1 of 2}
\label{tab:comparative_overview_symbolic_execution_tools}

    \centering
    \begin{tabular}{|c|c|c|c|c|}
        \hline
        \textbf{\makecell{Tool Name\\Reference}} &
        \textbf{\makecell{Primary\\ Category}} &
        \textbf{Input Required} & 
        \textbf{Core Technique(s)} & 
        \textbf{Key Features}
%        \textbf{Primary Use Case(s)} & 
%        \textbf{Constraint Solver(s)} & 
%        \textbf{Limitation(s)/Trade-offs} &
%        \textbf{Reference}
        \\
        \hline
        KLEE~\cite{cadar2008klee}& 
        Source-Based & 
        LLVM Bitcode & 
        \makecell{Symbolic Execution\\ Object-level COW\\ Constraint Independence\\ Counterexample Caching} & 
        \makecell{High-coverage test \\generation, Bug finding\\ Environment modeling} 
%        \makecell{Software Testing\\ Bug Detection\\ Source/Bitcode available} &
%        STP, Z3, etc. & 
%        Requires source/bitcode, Path explosion &
        \\ 
        \hline
        SymCC~\cite{Poeplau2020symbolic}  & 
        Source-Based & 
        LLVM Bitcode & 
        \makecell{Compile-time SE \\ Instrumentation, Concolic} &
        \makecell{High performance \\(avoids IR interp.)\\ Flexible solver integration} 
        \\
%        \makecell{Efficient test gen.\\Integration into build process} &
%        Z3 (default), others &
%        \makecell{Requires source/bitcode\\Newer tool/ecosystem} & 
        \hline
        \texttt{angr}~\cite{shoshitaishvili2016state}  & 
        Binary-Based & 
        \makecell{Binary\\(Multi-Arch)} & 
        \makecell{DSE, VSA,\\ CFG recovery,\\ Veritesting,\\Path prioritization} &
        \makecell{Python framework\\Flexible, Extensible\\Arch-agnostic\\Large community}
%        \makecell{Binary Analysis, \\Vuln. Research,\\ Malware Analysis,\\ RE, CTFs} &
%        Z3 (default), Claripy &
%        \makecell{Can be slow\\ Path explosion\\ Env. modeling complexity} &
        \\
        \hline
        Triton~\cite{saudel2015Triton}& 
        Binary-Based & 
        \makecell{Binary \\(Multi-Arch)} &
        \makecell{Concolic\\Dynamic Binary Analysis\\ Taint Tracking\\SMT Conversion} &
        \makecell{Python scripting\\ Runtime tracing\\ Taint analysis focus}
        \\
%        \makecell{Obfuscated/Malware analysis\\ RE, Dynamic analysis} &
%        Z3 (via SMT2-LIB) & 
%        \makecell{Relies on Intel Pin (mostly x86)\\ Input determinism assumption} &  \\
        \hline
        $S^2E$~\cite{chipounov2011S2E} & 
        \makecell{Full-System\\ Binary-Based} & 
        \makecell{Binary x86\\ Full System Image} & 
        \makecell{Selective Symbolic Execution\\ Virtualization (QEMU)\\DBT and Relaxed \\Consistency Models} & 
        \makecell{Full system analysis\\(incl. kernel/drivers)\\ In-vivo analysis\\No env. models needed\\Consistency trade-offs}
%        \makecell{Firmware analysis\\Driver testing/RE\\OS-level analysis\\Perf. analysis} &
%        KLEE solvers (STP, Z3) &
%        \makecell{Complex setup\\Performance overhead\\ State explosion} &
        \\
        \hline
        BINSEC/SE~\cite{david2016binsec}  &
        Binary-Based &
        \makecell{Binary \\(Multi-Arch)} &
        DSE & 
        Modular, Configurable
%        \makecell{Malware exploration\\ Reverse Engineering} &
%         &
%        Relies on PIN traces &
        \\
        \hline
        Driller~\cite{stephens2016driller}  &
        \makecell{Hybrid\\Binary-Based} &
        Binary &
        \makecell{Fuzzing (AFL) + \\Selective Concolic \\Execution (\texttt{angr})} &
        \makecell{Combines fuzzing breadth +\\ SE depth\\Finds deep bugs} 
        \\
%        Vulnerability Research (deep bugs) & 
%        angr solvers (Z3) &
%        \makecell{Complexity of coordinating fuzzer/SE\\ Relies on underlying tools} & \\
        \hline
    \end{tabular}

%\end{table}
\end{sidewaysfigure}
%
% Part 2 of 2 of the full page sideways figures
%
\begin{sidewaysfigure}[p]
\caption{Comparative Overview of Symbolic Execution Tools, Part 2 of 2}
\label{tab:comparative_overview_symbolic_execution_tools2}
    \centering
    \begin{tabular}{|c|c|c|c|}
        \hline
        \textbf{\makecell{Tool Name\\Reference}} &
%        \textbf{Primary Category} &
%        \textbf{Input Required} & 
%        \textbf{Core Technique(s)} & 
%        \textbf{Key Features} & 
        \textbf{Primary Use Case(s)} & 
        \textbf{Constraint Solver(s)} & 
        \textbf{\makecell{Limitation(s)\\Trade-offs}} \\
%        \textbf{Reference}
        \hline
        KLEE~\cite{cadar2008klee}& 
 %       Source-Based & 
 %       LLVM Bitcode & 
 %       \makecell{Symbolic Execution\\ Object-level COW\\ Constraint Independence\\ Counterexample Caching} & 
%        \makecell{High-coverage test generation\\ Bug finding\\ Environment modeling} & 
        \makecell{Software Testing\\ Bug Detection\\ Source/Bitcode available} &
        STP, Z3, etc. & 
        \makecell{Requires source/bitcode\\Path explosion} 
        \\ 
        \hline
        SymCC~\cite{Poeplau2020symbolic}  & 
%        Source-Based & 
%        LLVM Bitcode & 
%        \makecell{Compile-time SE \\ Instrumentation, Concolic} &
%        \makecell{High performance (avoids IR interp.)\\ Flexible solver integration} & 
        \makecell{Efficient test gen.\\Integration into build process} &
        Z3 (default), others &
        \makecell{Requires source/bitcode\\Newer tool/ecosystem}
        \\
        \hline
        \texttt{angr}~\cite{shoshitaishvili2016state}  & 
%        Binary-Based & 
%        \makecell{Binary\\(x86, ARM, MIPS, etc.)} & 
%        \makecell{DSE, VSA,\\ CFG recovery,\\ Veritesting,\\Path prioritization} &
%        \makecell{Python framework\\Flexible, Extensible\\Arch-agnostic\\Large community} & 
        \makecell{Binary Analysis, VR,\\ Malware Analysis,\\ RE, CTFs} &
        Z3 (default), Claripy &
        \makecell{Can be slow due to Path explosion\\ Env. modeling complexity}
        \\
        \hline
        Triton~\cite{saudel2015Triton}& 
%        Binary-Based & 
%        \makecell{Binary \\x86/x64 via Pin} &
%        \makecell{Concolic\\Dynamic Binary Analysis\\ Taint Tracking\\SMT Conversion} &
%        \makecell{Python scripting\\ Runtime tracing\\ Taint analysis focus} & 
        \makecell{Obfuscated/Malware analysis\\ RE, Dynamic analysis} &
        Z3 (via SMT2-LIB) & 
        \makecell{Relies on Intel Pin (mostly x86)\\ Input determinism assumption} 
        \\
        \hline
        $S^2E$~\cite{chipounov2011S2E} & 
%        Full-System / Binary-Based & 
%        \makecell{Binary (x86, ARM)\\ Full System Image} & 
%        \makecell{Selective Symbolic Execution\\ Virtualization (QEMU)\\DBT\\Relaxed Consistency Models} & 
%        \makecell{Full system analysis\\(incl. kernel/drivers)\\ In-vivo analysis\\No env. models needed\\Consistency trade-offs} & 
        \makecell{Firmware analysis\\Driver testing/RE\\OS-level Perf. analysis} &
        KLEE solvers (STP, Z3) &
        \makecell{Complex setup\\Performance overhead\\ State explosion}
        \\
        \hline
        BINSEC/SE~\cite{david2016binsec}  &
 %       Binary-Based &
 %       Binary (via PIN trace) &
 %       DSE & 
 %       Modular, Configurable & 
        \makecell{Malware exploration\\ Reverse Engineering} &
         (Z3, CVC4, Boolector) &
        Relies on PIN traces
        \\
        \hline
        Driller~\cite{stephens2016driller}  &
 %       Hybrid / Binary-Based &
 %       Binary &
 %       \makecell{Fuzzing (AFL) + \\Selective Concolic Execution (angr)} &
%        \makecell{Combines fuzzing breadth +\\ SE depth\\Finds deep bugs} &
        \makecell{Vulnerability Research\\ (deep bugs)} & 
        \texttt{angr} solvers (Z3) &
        \makecell{Complexity of coordinating fuzzer/SE\\ Relies on underlying tools} 
        \\
        \hline
    \end{tabular}

%\end{table}
\end{sidewaysfigure}

These selected tools represent a broad spectrum of symbolic execution approaches, from source-based test generation (KLEE, SymCC) to binary-focused analysis (\texttt{angr}, Triton, $S^2E$, Driller). Their differing methodologies, integrations with constraint solvers, and architectures highlight the flexibility and adaptability of symbolic execution to various analysis challenges. Subsequent sections will reference these tools as examples when discussing how symbolic execution addresses specific problems in areas like vulnerability research, firmware analysis, and obfuscated code analysis.

\section{Applications and Guidance Strategies in Symbolic Execution}
\label{sec:analysis-challenges-and-tool-evaluations}

This section surveys the practical application of symbolic execution across several challenging application domains. The tools presented in Section ~\ref{sec:tools-landscape} have diverse design and goals. As a result, they are better suited towards specific tasks. For instance, source-based tools like KLEE are ideal in a development environment, but ill-suited for analyzing malware, which is typically obfuscated and source code is not available.

A unifying theme emerges from the surveyed literature: the practical success of symbolic execution hinges on the ability to intelligently its inherent complexity. Pure, unguided exploration is often intractable for any non-trivial program, a fact established in Section ~\ref{subsec:common-challenges-in-symbolic-execution}. Consequently, the practical applications detailed in this section are the result of applying various strategies to make the analysis possible. These strategies can be broadly categorized into:
\begin{itemize}
    \item Scope Reduction, which limits the analysis to smaller manageable units of code.
    \item Guidance Heuristics, which intelligently steer the search to prioritize more promise execution paths.
\end{itemize}
Recently, it has become common to utilize hybrid analysis to implement the strategies above. Hybrid analysis combines symbolic execution with some another analysis technique (often fuzzing) to complement their strengths. This hybrid architecture is not a separate category of complexity management, but rather a powerful pattern for achieving scope reduction, by applying symbolic execution on targeted pieces of code, or implementing guidance heuristics.

Out of the categories outlines above, guidance heuristics represent a diverse and powerful set of techniques.  Table ~\ref{tab:guidance_strategies} presents a taxonomy of common guidance strategies that will serve as an analytical framework for the remainder of this section. 

%At some point cite ~\cite{sabbaghi2020systematic}

\begin{table}[htbp]
  \scriptsize
  \centering
  \renewcommand{\arraystretch}{0.3} % Increases vertical space between rows
  \begin{tabularx}{\textwidth}{| >{\raggedright\arraybackslash}X | >{\raggedright\arraybackslash}X | >{\raggedright\arraybackslash}X | >{\raggedright\arraybackslash}X |}
    \hline
    \textbf{Guidance Strategy} & \textbf{Guiding Principle} & \textbf{Example Paper(s)} & \textbf{Target Domain} \\ \hline
    Bug-Driven / Vulnerability-Oriented & Prioritize paths that are statistically or structurally more likely to contain known bug patterns (e.g., near sanitizer alerts or type-unsafe operations). & Saviour, UAFDetect, Vital & General Vulnerability Research \\ \hline
    Specification-Guided & Explore paths to check for conformance with, or violations of, a formal protocol specification (e.g., an RFC). & Wen et al., Asadian et al. & Protocol Conformance Testing \\ \hline
    Model-Guided & Learn a state machine or model of the system’s behavior first, then use that model to guide exploration towards interesting or uncovered states. & MACE, PISE & Protocol Inference \& Analysis \\ \hline
    Goal-Directed & Focus exploration exclusively on finding a path to a specific target state or program location, pruning all other paths. & Jetset & Firmware Re-hosting, Targeted Analysis \\ \hline
    Invalidity-Guided & Learn correct system behavior by analyzing and reasoning about inputs that cause the system to enter an invalid state (e.g., a crash or stall). & uEmu & Firmware Peripheral Modeling \\ \hline
  \end{tabularx}
  \caption{Guidance Strategies in Software Analysis}
  \label{tab:guidance_strategies}
\end{table}

\subsection{Program Verification}
\label{subsec:program-verification}

A key goal in software development is verifying correctness. For simple systems, this is straightforward, however, as software complexity increases, so does the task of validating correctness across all possible inputs. Symbolic execution enables formal verification by exhaustively exploring program paths and checking properties like safety, liveness, or adherence to protocols. The works surveyed here provide examples in all three broad categories outlined Table ~\ref{tab:guidance_strategies}.

As noted earlier scope reduction is one strategy to enable scalable symbolic execution. For example under-constrained symbolic execution~\cite{engler2007under} which analyzes individual functions in isolation. Ramos and Engler~\cite{ramos2015under} apply this technique in UC-KLEE, a framework that bypasses the costly analysis from a program's \texttt{main} function to test individual functions in isolation. Hybrid analysis is another powerful option which combine the analysis prowess of symbolic execution with a faster technique. Map2Check~\cite{rocha2020map2check}, for example, adopts this approach by integrating fuzzing to quickly find shallow bugs with symbolic execution used for deeper path exploration. Finally, many approaches rely on guidance heuristics. In the specialized domain of hardware verification, SEIF provides a clear example of a model-guided solution. It addresses information flow in RTL Verilog designs by building a static signal connectivity graph, which then acts as a guide for the symbolic execution engine~\cite{ryan2023sylvia}. SEIF prunes paths that are logically or architecturally impossible.

These approaches, while powerful, highlight persistent challenges. Techniques like under-constrained symbolic execution can suffer from false positives due to missing context or unresolved external calls, while hybrid methods still struggle with complex data structure modeling. Domain-specific approaches like SEIF are effective but may require bounding analysis depth, potentially missing subtle bugs. Future work in this area will likely focus on automatically inferring function preconditions and invariants to reduce false positives and more tightly integrating diverse analysis techniques to improve both speed and precision.

\subsection{Vulnerability Research}
\label{subsec:vulnerability-research}

Vulnerability research is a critical area of study that interests security practitioners, white hat hackers, and developers alike. Identifying vulnerabilities in large, complex software systems is challenging due to code size, execution complexity, and diverse behaviors. Additionally, it can be difficult to verify that a vulnerability identified in an isolated function is truly reachable from an external entry point.

The strength of symbolic execution lies in its ability to reason about a program's internal behavior and automatically generate concrete inputs that trigger specific bugs. This greatly simplifies the task of creating proof-of-concept exploits, streamlining the verification and reproduction of vulnerabilities. However, given the vast search space of modern software, a dominant theme in this area is the use of bug-driven guidance to better explore the program. This strategy focuses the symbolic execution engine on paths that are more likely to contain bugs. 

This section surveys various approaches that tailor this bug-driven strategy to specific bug classes, discussing their strengths and weaknesses. 

\paragraph{Memory Corruption Vulnerabilities:}

Memory corruption vulnerabilities represent the majority of vulnerabilities discovered in software. Despite the software community's comprehensive understanding of various memory corruption vulnerability classes (e.g., buffer overflows, use-after-free (UAF) vulnerabilities), these flaws remain challenging to detect reliably. Their triggers often depend on intricate program states and specific input sequences, making them challenging to detect. Symbolic execution provides an approach to reliably detect these code violations. 

Many tools focusing on bug finding leverage a hybrid analysis architecture while implementing a bug-driven guidance heuristic. SAVIOR~\cite{chen2020savior} represents such an example. It uses UbSan~\cite{UBSanClang} to label suspicious code regions, guiding the fuzzer toward potentially vulnerable paths, and then uses selective symbolic execution to validate errors and generate concrete test cases. UAFDetect~\cite{huang2023targeted} uses a bug-driven guidance heuristic approach to constrain symbolic execution and operates in two phases. First, it statically identifies potential UAF code regions. Second, it employs symbolic execution to perform dynamic typestate analysis~\cite{strom1986typestate} to track the lifecycle of pointers (e.g., Init, Allocated, Deallocated). UAFDetect prunes paths that are unlikely to cause UAF violations. Vital~\cite{tu2024vital} also utilized a bug-driven approach guiding symbolic execution towards paths containing a high number type-unsafe pointers. The key insight here is that paths with a greater number of type-unsafe pointers are more likely to contain vulnerabilities. VITAL uses Monte Carlo Tree Search (MCTS) to proactively guide KLEE toward vulnerability-rich paths, prioritizing paths in the symbolic execution tree that contain larger numbers of unsafe pointers. 

To tackle binary-only targets, researchers often turn to angr. UbSym~\cite{baradaran2023unit} employs scope reduction introducing a unit-based methodology that addresses scalability challenges differently from the source-code approaches. Instead of applying symbolic execution to whole programs, UbSym breaks programs into smaller units and performs symbolic analysis on individual units that are statically identified as potentially containing vulnerabilities based on defined memory corruption vulnerability specifications. When vulnerabilities are detected at the unit level, UbSym utilizes machine learning techniques, specifically the TAR3 learning algorithm~\cite{menzies2003data}, a machine learning heuristic for feature ranking, to predict which paths will trigger program-level vulnerabilities.

Collectively, these studies demonstrate a clear trend: guided symbolic execution is essential for efficiently targeting memory corruption vulnerabilities. The strategies for guiding symbolic execution vary significantly based on both the vulnerability class being targeted and the analysis scope. SAVIOR, UAFDetect, and Vital all utilize a two-pass system. First, features are discovered through static analysis, then guiding symbolic execution along paths deemed interesting. In contrast, UbSym focuses on structural properties, decomposing the program and using ML predictions based on unit-level behavior.  
% this is a good insight - symbolic execution must be guided in order to be feasible
Furthermore, these approaches differ in scope and tooling. SAVIOR, UAFDetect, and Vital operate at the whole-program level and leverage KLEE as their symbolic execution engine. This makes them more suited for analysis when source code is available. UbSym's use of \texttt{angr} and unit-level decomposition showcases an approach for tackling binary-only targets and managing scalability by analyzing smaller code chunks. These complementary approaches highlight the diversity of strategies available for addressing the path explosion problem in symbolic execution while maintaining focus on vulnerability detection.

\paragraph{Concurrency Vulnerabilities:} 
Analyzing multi-threaded programs is a challenge for even human programmers. In symbolic execution, a key challenge in analyzing multi-threaded programs is managing the dual nondeterminism of both symbolic inputs and thread scheduling, both of which exacerbate the path explosion problem. Within concurrency analysis, research has tackled both the correctness of thread interleaving and the raw scalability of the analysis. Research in this space shift the challenge from "which of these millions of paths is most interesting" to "how do we manage the explosion of thread interleavings and shared state?" The surveyed works generally fall into one of two categories: Managing Interleaving Explosion and Providing Foundational Support and Scalability.

To address interleaving explosions, Schemmel et al.,~\cite{schemmel2020symbolic} integrate symbolic execution into a partial-order reduction~\cite{abdulla2014optimal} (POR). They developed a KLEE-based prototype that constructs an unfolding of the program that captures concurrent behaviors while merging equivalent execution prefixes. To prune redundant paths, they employ cutoff events, detecting when a state has already been reached via a shorter path, and restrict scheduling decisions to synchronization primitives (mutex, condition variables), also reporting data races as errors. While this technique has proved effective at identifying previously unknown vulnerabilities in Memcached~\cite{MemcachedWebsite}, it focuses on reducing interleaving explosion rather than providing comprehensive threading models. As a result, it may still face scalability issues with very large programs. To address the separate challenge of scalability, researchers have explored distributing the symbolic workload itself. Bucur et al.,~\cite{bucur2011parallel} parallelize symbolic execution across a commodity cluster using a dynamic load-balancing scheme. The Cloud9 system, built on top of KLEE, distributes states among multiple nodes and provides a comprehensive symbolic model of the POSIX interface, including robust support for threads, networking, and I/O. 

To provide better foundational support for analyzing multi-threaded programs, two works provide prominent examples. Notable efforts from Vishnyakov et al.'s Symbolic DynamoRIO (Sydr)~\cite{vishnyakov2020sydr}, which leverages the Triton symbolic engine to analyze multi-threaded binaries. Sydr manages separate symbolic states for each thread and handles context switches but, as the authors note, it currently does not influence thread scheduling, a feature planned for future work. Similarly, Niskov et al.~\cite{niskovenhancing} introduced enhancements to the $S^2E$ platform. Their contributions include enabling $S^2E$ to support multiple virtual cores and implementing a plugin that functions as a data race checker for multi-threaded programs. This race checker's design is inspired by the $DJIT^+$ algorithm~\cite{pozniansky2003efficient}, aiming to detect concurrency defects within the symbolic execution framework. These developments signify important strides in extending the capabilities of binary symbolic execution to the complex domain of multi-threaded software analysis.

The primary limitations in this area remain performance and modeling complexity. Even with techniques like POR, memory usage can become prohibitive for long-running programs, and not all tools can yet influence thread scheduling to explore specific interleaving. Future improvements will likely focus on more advanced heuristics for path pruning and richer modeling of both synchronization primitives and external library calls to increase precision and scalability.

\paragraph{Specialized Vulnerability Domains:} 
The previous paragraphs covered research in well-known application domains. However, symbolic execution has applications in a wide variety of specialized areas. While we cannot discuss all of the different domains, we highlight research we found most interesting. Specifically, we examine how symbolic execution has been applied to specialized domains, focusing on smart contracts and microarchitecural side-channel attacks.

In the smart contract space, we highlight two prominent studies, both taking a learning-based approach. He et al.~\cite{he2019learning} employ a hybrid approach and introduce Imitation Learning Fuzzer (ILF) where a fuzzer learns to imitate the behavior of a symbolic execution 'expert'~\footnote{ILF utilizes the VerX symbolic execution engine}, aiming to combine the strengths of both techniques. The core idea is for the fuzzer to learn an effective fuzzing policy for generating input sequences by observing a symbolic execution 'expert'. Because symbolic execution of entire smart contracts is expensive even for small contracts, the expert symbolically executes a individual transactions and concretizes inputs to optimize coverage. The symbolic execution engine generates inputs leading to maximum coverage or indicates when no input would improve coverage. This enables the fuzzer to achieve better coverage than existing smart contract fuzzers. ILF's primary goal is to detect vulnerabilities in Ethereum~\cite{wood2014ethereum} smart contracts~\footnote{The specific vulnerabilities are Locking, Leaking, Suicidal, Block Dependency, Unhandled Exception, and Controlled Delegatecall} using specific detectors for each vulnerability class. MythrilQL~\cite{wang2023reinforcement} is an example of bug-driven guidance. It leverages Q-learning as a mechanism to guide the symbolic execution engine's path selection towards vulnerable paths. Additionally, they implement an incentive based path pruning strategy to eliminate redundant or useless paths. Finally, for constraint solving efficiency, they predict solution time to determine whether a constraints should be passed to the solver.

Another compelling research direction utilizes symbolic execution to identify Spectre vulnerabilities~\cite{kocher2020spectre}. These vulnerabilities differ from traditional memory safety bugs because they only manifest during speculative execution. That is when processors execute instructions based on predictions that may prove incorrect. Attacks against these vulnerabilities exploit side effects resident by these executions even if the processor reverts its state when predictions fail. The challenge for detection tools is modeling both regular execution paths and speculative (transient) paths simultaneously while managing exponential explosion of possible execution traces. This creates a unique state explosion problem, similar to the interleaving explosion in concurrent programs. To tackle this, Daniel et al.'s~\cite{daniel2021hunting} BINSEC/HAUNTED utilize relational symbolic execution~\cite{palikareva2016shadow, farina2019relational} (RelSE)~\footnote{Relational symbolic execution is a technique that analyzes pairs of execution traces simultaneously.} to tackle the challenge of detecting Spectre vulnerabilities~\cite{kocher2020spectre}. This allows them to analyze both regular and transient execution paths in a single run, enabling efficient reasoning about side-channel leaks caused by speculative misprediction. Haunted RelSE represents transient executions alongside regular executions without explicitly exploring all speculative paths.

While these specialized approaches show great promise, their focus is also their limitation. Learning-based tools like ILF may struggle with smart contracts that have unique logic that is not present in the training data. Similarly, while BINSEC/HAUNTED scales better than explicit speculative exploration, it remains infeasible for very large binaries and is currently limited to detecting specific Spectre variants (PHTand STL). Future work could refine these specialized heuristics and extend them to cover a broader range of behaviors of attack classes.

\subsection{Obfuscated and Malicious Code Analysis}
\label{subsec:obfuscated-malicious-code}

Obfuscation is often used to hinder reverse engineering and make static analysis difficult by transforming code logic into more complex or misleading forms. Common techniques include control-flow flattening, encryption of constants, and self-modifying code, all of which complicate attempts to reason about program behavior. Malware authors routinely exploit such obfuscation to evade detection, creating an ongoing arms race where security researchers strive to uncover malicious intent and shared code patterns across evolving malware variants.

Symbolic execution offers a powerful countermeasure by systematically exploring paths within obfuscated binaries and reasoning about hidden logic. That is, symbolic execution can reveal malicious routines, identify hidden command structures, and generate meaningful inputs to drive deeper analysis. However, symbolic execution still faces significant hurdles in practice: large search spaces, runtime-based anti-analysis, and environment-specific dependencies can all limit its effectiveness. To overcome these hurdles, researchers have employed a variety of strategies, including goal-directed guidance to focus on malicious behaviors, scope reduction to make large-scale analysis feasible, hybrid analysis to combine complementary techniques, and specialized methods like invalidity-guided analysis to defeat specific obfuscation schemes.

Several approaches use symbolic execution to extract semantic features for malware classification and analysis. For instance, SEMA~\cite{bertrand2023SEMA} utilize a goal-directed guidance strategy to build System Call Dependency Graphs (SCDGs) that serve as behavioral signatures. The goal is not necessarily a specific line of code, but rather to generate a behavioral signature. To alleviate path explosion, they implement a custom breadth-first search strategy~\cite{vanouytsel2022malwareanalysissymbolicexecution} that prioritizes interesting paths for exploration. SEMA extends \texttt{angr} with specialized strategies to create representative signatures based on SCDGs to feed machine learning models for classification. Vouvoutsis et al.~\cite{vouvoutsis2025beyond} take a different approach, applying scope reduction at the dataset level. They developed a detection pipeline that balances accuracy and automated detection of evasive malware. Their pipeline clusters malware using TLSH, performs symbolic execution on representatives from each cluster using \texttt{angr} to gather API call features, applies unification to the extract API features, and finally feeds these features to machine learning classifiers. Beyond classification, Botacin et al.~\cite{botacin2021malware} introduce Malverse, which uses a goal-directed approach to automatically identify execution paths hidden behind evasive conditions and automatically patch the binary. Malverse uses a Bayesian model as the heuristic to steer symbolic execution toward potentially malicious paths.  Malverse uses symbolic execution to discover function inputs and returns that trigger malicious behaviors, the malware is then patched with values satisfying the conditions forcing the malware to execute its malicious behavior.

Symbolic execution has also been employed to automatically deobfuscate code. Salwan et al.~\cite{Salwan2018symbolic} implement a unique hybrid approach that combines symbolic execution with taint analysis to automatically deobfuscate virtualized code. Their approach, built on Triton, uses taint analysis to isolate the pertinent instructions related to the original program logic, and then uses symbolic execution to reconstruct their behavior. Another common obfuscation technique is an opaque predicate. Bardin et al.~\cite{bardin2017backward} take an invalidity-guided approach. The authors introduce a technique, Backward-Bounded Dynamic Symbolic Execution (BB-DSE), to identify opaque predicates by proving their branches are unreachable. BB-DSE works backward from a suspicious region of code to determine if they can ever be satisfied allowing the obfuscation artifacts to be eliminated.

These approaches reveal several methods in which symbolic execution provides leverage against obfuscated and evasive malware. While powerful, these applications highlight the persistent challenges of scalability and environment modeling. Strategies to cope with these challenges include reducing the analysis scope via clustering (Vouvoutsis et al.), employing heuristics and targeted analysis (SEMA, MalVerse, BB-DSE), and focusing on specific, constrained problems (Salwan et al).

\subsection{Emulation and Firmware Analysis}
\label{subsec:emulation-and-firmware-analysis}
Embedded systems are becoming increasingly popular. Unfortunately, their increased popularity and widespread use has not resulted in better software hygiene when developing for embedded systems. Because these systems are often constrained in both compute power and storage, developers want to get the absolute most out of the hardware on these systems. As a result, security is often overlooked for the sake of performance.

Firmware analysis presents unique challenges due to its dependence on hardware for execution. Tools like QEMU\footnote{\url{https://www.qemu.org/}} 
and Unicorn \footnote{\url{https://github.com/unicorn-engine/unicorn}} allow firmware emulation, but replicating behavior of hardware peripherals often requires detailed hardware specifications that are not always available. The concept of firmware re-hosting was usually done in ad-hoc environments for smaller projects. However, the rise of the use of embedded devices has brought scrutiny to the field of firmware re-hosting. Fasano et al~\cite{fasano2021enabling} described the challenges in re-hosting embedded devices. In order to properly emulate firmware, you often need a significant amount of information about the hardware (e.g., memory maps, peripheral register behaviors) which is not always readily available. Symbolic execution can bridge this gap by learning peripheral responses dynamically, reducing the reliance on hardware models. Research in the area of firmware emulation has diverged into two main camps: high-fidelity analysis using hardware-in-the-loop (HIL), and hardware-free analysis that infers peripheral behavior using strategies like hybrid analysis and various forms of guidance.

HIL approaches offer the best fidelity. By incorporating physical hardware, there is no need to speculate peripheral responses or firmware behavior. Avatar2~\cite{muench2018avatar} stands out not as an analysis technique, but rather a mature orchestration platform. It addresses the interoperability problem between analysis tools by enabling coordination among emulators (QEMU, PANDA), debuggers (GDB, OpenOCD), symbolic execution engines (\texttt{angr}), and physical hardware. While \texttt{angr} was used in their original implementation, Avatar2's modular design allows many different configurations. Another HIL technique is CO3~\cite{liu2024co3} which utilizes a hybrid approach. CO3 brings hybrid (concolic) execution to resource-constrained MCUs by offloading the complex analysis to a powerful workstation while executing on the real device.

In contrast, several systems use symbolic execution to enable firmware analysis without physical hardware by modeling or inferring peripheral behavior. These approaches differ in their specific goals and methodologies. Despite a lack of hardware, a hybrid approach is still feasible through emulation. For instance, Fuzzware~\cite{scharnowski2022fuzzware} takes a hybrid approach which uses \texttt{angr} to learn precise Memory Mapped Input Output (MMIO) models that guide fuzzing to more meaningful inputs. Specifically, Fuzzware uses \texttt{angr} to determine precise models of how the firmware reads from MMIO registers allowing Fuzzware to build abstract models providing meaningful choices to the fuzzer. For instance, if \texttt{angr} classifies a peripheral interaction as constant, the fuzzer consistently produces the same value for that peripheral access. uEmu~\cite{zhou2021automatic} takes a different approach. It uses invalidity-guided symbolic execution with $S^2E$ to infer correct register responses by learning from states that cause the firmware to crash or stall. The peripheral responses are stored in a knowledge base which can be used to subsequently run the firmware in an emulator for fuzzing. For targeted re-hosting scenarios, Jetset~\cite{johnson2021jetset},implemented using \texttt{angr} and QEMU, uses goal-directed symbolic execution to find only the specific I/O constraints necessary to boost a firmware image to a target address. These methods demonstrate a trade-off between the high fidelity of HIL systems and the broader applicability and scalability of peripheral inference techniques.

Firmware analysis is challenging due to how tightly coupled the behavior is tied to specific hardware peripherals, which are often undocumented or hard to accurately emulate. While Avatar2 and CO3 aim for high fidelity by integrating hardware in their analysis, uEmu, Jetset, and Fuzzware take a different approach by utilizing symbolic execution to infer peripheral behavior. Still, the approaches to infer peripheral behavior differ significantly. uEmu uses invalidity guidance to infer correct register responses. Jetset uses goal-directed symbolic execution to find constraints just sufficient to reach a target state for rehosting. Finally, Fuzzware uses symbolic execution to learn precise, compact models of MMIO behavior to optimize fuzzing. Each approach has its own tradeoffs and goals. However, the key insight across all peripheral modeling approaches is that emulating sufficient peripheral behavior to enable firmware execution and testing does not require perfect accuracy.

\subsection{Protocol Inference and State Analysis}
\label{subsec:protocol-inference-and-state-analysis}

Effective communication requires protocols that establish an agreement between the communicating parties on how communication is to proceed~\cite{tanenbaum2011computer}. While formal methods can verify protocol designs using tools such as CPSA~\cite{CPSA4Manual}, ensuring that a software implementation correctly adheres to its specification is a distinct and critical challenge. Deviations from specifications can introduce vulnerabilities, potentially subverting the protocol's intended security guarantees. For example, implementation vulnerabilities such as the Heartbleed vulnerability~\cite{cve20140160,openssl2014heartbleed}, the POODLE attack~\cite{cve20143566}, and a recent OpenSSL Vulnerability~\cite{CVE-2024-12797} arise from deviations or errors in protocol implementations. This challenge is heightened when analyzing proprietary or malicious protocols (e.g., malware command and Control), where the specification is unknown and must be reverse engineered.

Symbolic execution is a powerful technique for this domain, but the stateful nature of protocol can quickly lead to path explosion. To manage this complexity, researchers primarily employ two powerful guidance strategies: model-guided analysis, where a state machine is learned from the implementation, and specification-guided analysis, which uses a formal specification, such as an RFC, to direct the search for bugs.

A dominant strategy is model-guided symbolic execution, where a state machine is first learned from the implementation and then used to steer analysis. L*~\cite{angluin1987learning} is the primary algorithm used to infer a protocol. MACE~\cite{cho2011mace}, PISE~\cite{Marcovich2022PISE}, and Wen et al.~\cite{wen2017model} employ this strategy. MACE iteratively infers a finite-state protocol model from the application's input/output behavior, which is then used to guide concolic execution using DART~\cite{godefroid2005dart}. PISE uses symbolic execution with \texttt{angr} to guide message exchanges and pair it with an extended L* automata-learning algorithm. By instrumenting send/receive functions and systematically constraining message bytes, PISE discovers valid protocol states and message formats through incremental collision resolution and alphabet refinement. While we have primarily used our taxonomy to label research into one of the guidance strategies, we note that some systems may employ a blend of these strategies. Wen et al. combine both approaches: they start with an RFC-defined message format (spec guidance) to build an initial model, then use L* to infer a state machine (model-guided). With this, they are able to utilize L* to generate a finite-state machine which guides the symbolic execution engine, $S^2E$, to less traversed paths. Asadian et al.~\cite{asadian2022applying} follow a purely specification-driven approach. First, they translate the protocol's RFC message and state rules into logical assertions for KLEE. Then KLEE explores those assertions to uncover implementation paths that conflict with the spec. To demonstrate the effectiveness of their approach, the authors applied it to the Datagram Trannsport Layer Security (DTLS) protocol and successfully reproduced CVE-2014-0195. The primary drawback is manually extracting and formalizing an RFC's details is labor-intensive and potentially error-prone. Another interesting approach is from Sun et al.~\cite{sun2022spenny} who introduce Spenny, which uses a "field coverage" metric to guide the analysis of proprietary Industrial Control System (ICS) protocols from firmware binaries. A fundamental challenge in this domain is handling cryptography; Vanhoef and Piessens~\cite{vanhoef2018symbolic} tackle this by simulating cryptographic primitives under the Dolev-Yao model~\cite{dolev1983security}, enabling analysis of security protocol implementations without getting stalled by complex SMT constraints.

As we observed in Section~\ref{subsec:vulnerability-research}, guiding symbolic execution is crucial especially when dealing with stateful protocols. The guidance provided varies from inferred models, specifications, or domain-specific heuristics. As with all other challenges, path/state explosion remains a concern even with guidance. Furthermore, handling complex environment interaction (network responses, timing) complicates analysis.

\section{Challenges and Future Directions}
\label{sec:challenges-and-future-directions}
While symbolic execution has matured into a practical analysis technique, its full potential is constrained by foundational challenges that continue to drive research. As discussed in Section~\ref{subsec:common-challenges-in-symbolic-execution}, issues such as path explosion, the computational cost of constraint solving, and the complexities of environment modeling remain significant hurdles. While much progress has been made, future work will likely focus not only on refining existing solutions but also on pioneering new approaches that integrate symbolic execution more deeply into the software life-cycle and apply it to increasingly sophisticated problem domains. This section outlines several promising research directions, moving from core technique enhancements to domain-specific applications. 

\subsection{Adapting Symbolic Execution to Real-Time Operating Systems}
Throughout this survey, we have presented the plethora of applications in which symbolic execution has been used. The application of symbolic execution in analyzing real-time operating systems (RTOS) remains underrepresented. RTOSes are widely deployed in safety and mission-critical applications including automotive, aerospace, and medical industry. Given their key role in critical systems, ensuring their implementation are correct and secure is essential. RTOSes pose unique analysis challenges. because timing guarantees are fundamental to their correct operation Modeling this in a symbolic execution environment requires explicitly representing time with symbolic variables. This would enable the engine to reason about deadlines and worst-case execution times (WCET). The challenge lies in determining which timing-related constraints should be modeled while balancing precision with practicality.

Furthermore, RTOSes typically leverage multitasking with priority-based scheduling, preemptions, and interrupts, all of which significantly hinder path exploration. As with the analysis of multi-threaded systems, symbolic execution must consider various ways tasks can interleave under preemption and time constraints. As a result, the path explosion problem is amplified. Partial-order reduction has proved effective at managing thread interleavings, however, it would like need to be adapted to fit this specific execution environment.

\subsection{Automated Characterization of Evasive Triggers in Malware}
Symbolic execution is often applied to extract features for classification, described in 
%Section ~\ref{subsubsec:relevant-research-obfuscated}. 
Section ~\ref{subsec:obfuscated-malicious-code}. 
% \gripe{is that the right section?}
However, a significant opportunity exists to move beyond classification and toward characterization, that is, understanding precisely how a malware specimen avoid detection. Symbolic execution could be used to automate this characterization to pinpoint the exact environmental conditions a malware sample checks. Instead of just bypassing a check, the research would focus on using the path constraints generated by the symbolic execution to produce a human-readable "trigger signature." For example, the analysis would not just find an input to bypass an anti-VM check; it would produce the very predicate the malware is solving for: (result\_of\_cpuid\_vendor\_string == "VMWareVMWare"). This research would address several open questions:
\begin{itemize}
    \item Can symbolic constraints generated from analyzing anti-analysis code be automatically simplified into concise, semantic predicates representing the evasion check?
    \item How can this technique scale to handle complex, multi-stage checks or environmental interactions (e.g., checking for specific registry keys, file paths, or running processes)?
    \item Could the resulting database of trigger signatures be used to automatically classify evasion techniques across different malware families, providing a deeper understanding of the attacker's toolkit?
\end{itemize}

\subsection{Analyzing Type-Safe Languages}
The increasing adoption of type-safe languages like Rust and Go is shifting the focus of vulnerability research from detecting memory corruption bugs to identifying logic errors and concurrency issues. This shift necessitates an evolution in how symbolic execution is applied. While significant research by Schemmel et al.~\cite{schemmel2020symbolic} and others has advanced the analysis of traditional thread-based concurrency in C programs, the structured concurrency models of modern languages like Rust and Go present new and distinct challenges. This future work would build upon foundational concepts like partial-order reduction but would require re-imagining them for these new paradigms.

For a language like Rust, a primary research thrust is the formal verification of unsafe blocks. Symbolic execution is an ideal tool to formally analyze these small, critical sections of code where the language's safety guarantees are manually suspended. Furthermore, with memory safety largely guaranteed in safe code, symbolic execution can focus on program correctness. This leads to questions such as: can symbolic execution find inputs that trigger denial-of-service conditions by causing a panic? Or can it verify higher-level semantic properties, such as "this function should never return an error if the input is positive"? While Rust's ownership model prevents data races at compile time, it does not prevent all concurrency bugs, such as deadlocks or logical race conditions. Symbolic execution could explore thread inter-leavings with Rust's concurrency primitives to find these higher-level bugs.

Similarly, a language like Go introduces its own unique modeling challenges, particularly concerning its garbage collector. Since the symbolic state of memory can be altered at any point by the garbage collector, this raises a critical research question: how can the effects of a garbage collector be soundly modeled within a symbolic execution engine without sacrificing performance, or can its effects be safely abstracted away?

\subsection{LLM-Assisted Symbolic Execution}
The rise of large language models (LLMs) presents an opportunity to blend these two technologies. While research in LLMs is still in its infancy, several promising directions have begun to emerge~\cite{wang2024python, li2025large, chen2025numscout, chen2024llm}. LLMs could be applied to address the primary challenges in symbolic execution: path explosion, constraint solving, and environment modeling.
\begin{itemize}
    \item Path Explosion: To manage potentially infinite state spaces, LLMs can assist in both the selection of paths and pruning of states. A hybrid system using an LLM can analyze a codebase to identify critical sections and guide symbolic execution towards those paths, avoiding those that are less relevant. This approach helps prioritize paths based on the user's goal and prune paths that do not align with those goals.
    \item Constraint Solving: LLMs can also be used to lessen the high computational cost of the constraint solving process. There are a few ways in which LLMs can be applied here. First, as a means of supporting new languages that are difficult to model. For example, utilizing an LLM to support the list data type in Python~\cite{wang2024python}. Alternatively, LLMs might be used as either a constraint solver for simpler queries, relieving the more expensive solver of that work, or as a means to simplify constraints before passing them onto a more robust solver.
    \item Environment Modeling: LLMs can provide great utility in environment modeling. Instead of requiring developers to manually write models for complex system calls or library functions, an LLM can be used to infer and predict the behavior of these external interactions~\cite{chen2024llm}. In this scenario, the LLM can act as the environment with which the symbolic execution engine interacts, improving the fidelity of the analysis.
\end{itemize}
A primary concern regarding using LLMs in symbolic execution is managing the probabilistic nature of LLMs.  In addition, working with LLMs could require more fine-tuned models specifically designed for the domain to ensure maximum fidelity and accuracy.

In addition to these directions, new application domains continue to emerge. The rise of IoT devices, automotive software, and even quantum computing platforms expands the frontier for symbolic execution, creating a need for multi-architecture, domain-specific solutions. While these challenges may not have immediate solutions, ongoing research seeks to adapt symbolic execution frameworks and constraint solvers to handle unconventional architectures and computation models.

Overall, the future of symbolic execution lies in making the technique more scalable, more versatile, and more accessible. This could be in the form of utilizing guidance strategies more effectively more robust environment modeling. By building on established suggestions from the literature and drawing on insights from research efforts in program analysis, automated reasoning, and domain-specific tool development, the community can continue to refine symbolic execution as a critical tool for understanding and securing complex software systems.

\section{Conclusion}
\label{sec:conclusion}
Symbolic execution continues to be a critical tool in various application domains. Our survey demonstrates that despite foundational issues such as path explosion, complex constraint solving, and accurate environment modeling (Section~\ref{subsec:common-challenges-in-symbolic-execution}), researchers have employed sophisticated guidance strategies and heuristics to mitigate these issues and broaden the applicability of symbolic execution.

Combining fuzzing with symbolic execution has shown promise in addressing the weaknesses of both techniques. Fuzzing often can "stall" and fail to discover new and interesting code paths. Hybrid approaches, exemplified in tools like Driller~\cite{stephens2016driller} and guided techniques like SAVIOR~\cite{chen2020savior}, capitalize on fuzzing's speed for broad exploration while using symbolic execution's analytical power to bypass complex checks. Future work could focus on enabling richer feedback between the fuzzer and symbolic engine. This might involve leveraging detailed constraint metrics or incorporating lightweight static analyses to more intelligently guide the exploration process.

In vulnerability research, guided symbolic execution has proven effective for memory corruption issues. Researchers have utilized heuristics based on runtime events ~\cite{chen2020savior, huang2023targeted}, static properties ~\cite{tu2024vital}, or structural decomposition ~\cite{baradaran2023unit}. A natural progression is to extend these guided techniques to detect other classes of bugs such as type confusion and logic errors at both the binary and source levels. 

The challenges posed by analyzing malware and obfuscated code continue to be of great interest. We've seen symbolic execution be used as another method to extract features for malware classification. Future directions in this area include developing techniques to systematically explore malware binaries and identify hidden checks the malware performs before triggering the malware. This could be taken further to use constraint solving to identify the exact check the malware performs (e.g., "Is the result of CPUID vendor string 'VMWare?'"). This would provide an automated identification and characterization of these evasion triggers directly from the binary code which could be used to classify different evasion techniques based on the patterns observed in their symbolic constraints.

In firmware and protocol analysis, the need to balance fidelity with scalability remains a challenge. Symbolic execution has been utilized to in hardware-software co-execution frameworks~\cite{liu2024co3, muench2018avatar}, for inferring peripheral behavior to enabled emulation or guided fuzzing~\cite{davidson2013FIE, zhou2021automatic, johnson2021jetset, scharnowski2022fuzzware}, and for protocol inference or conformance testing using learned models or specifications~\cite{cho2011mace, wen2017model, Marcovich2022PISE, sun2022spenny, asadian2022applying}. Future work could focus on improving the fidelity/scalability or achieving more complete automated inference of complex or encrypted protocols.

Overall, the future of symbolic execution likely lies in its intelligent integration with other techniques, fuzzing, static analysis, machine learning, and its specialization for challenging domains like firmware, protocols, and concurrent systems. These advances are essential as software systems grow ever more complex, ensuring that both vulnerabilities are detected early and that our defenses remain robust in the face of evolving threats.
%\nocite{*}
%\bibliographystyle{IEEEannot}
\printbibliography
\end{document}